\documentclass[thmsa]{article}
\usepackage{sw20lart}

\input tcilatex
\QQQ{Language}{
British English
}

\begin{document}

\section{Introduction}

We have recently derived generalized spin probability amplitudes, states and
operators and their eigenvectors [1-4]. The new quantities are more
generalized than the quantities considered most generalized hitherto in the
literature. In this paper, we show how these new results are related to the
standard results, and how the new results are used to calculate the
quantities of usual interest.

This paper is organized as outlined below. In Section $2$, we review the
standard interpretation of spin, so as to prepare the way for a comparison
with the interpretation arising from the new results. Section $3$ is devoted
to an exposition of the generalized interpretation of spin. We explain the
connection between probability amplitudes for spin projection measurements
and spin states in Section $3.1$. We give explicit forms for the spin states
in this section. In Section $3.2$, we give explicit formulas for the spin
operators and their eigenvectors. We then explain the connection between
spin states and the eigenvectors of spin operators in Section $3.3$.

Section $4$ is dedicated to the treatment of spin-projection expectation
values. Thus, the contrast between the standard and the generalized
interpretation of these expectation values is drawn in Section $4.1$. The
generalized results are presented in Section $4.2$. Section $4.3$ is a
summary of the various possibilities that arise when we compute these
expectation values by the new methods. The relation between these formulas
and the standard ones is clarified in Section $4.4$. The paper closes with a
discussion and conclusion in Section 5.

\section{Standard Interpretation of Spin}

We now briefly review the standard interpretation of spin. We shall
illustrate our discussion by means of the spin-$1/2$ case. The general
results will be valid for other values of spin.

Let the spin be measured in units of $\hbar /2.$ Therefore, the $z$
component of spin is given by the Pauli matrix

\begin{equation}
\lbrack \sigma _z]=\left( 
\begin{array}{cc}
1 & 0 \\ 
0 & -1
\end{array}
\right) ,  \label{one}
\end{equation}
whose eigenvectors are

\begin{equation}
\lbrack \xi _{+}]=\left( 
\begin{array}{c}
a_{+} \\ 
b_{+}
\end{array}
\right) =\left( 
\begin{array}{c}
1 \\ 
0
\end{array}
\right) \text{ and }[\xi _{-}]=\left( 
\begin{array}{c}
a_{-} \\ 
b_{-}
\end{array}
\right) =\left( 
\begin{array}{c}
0 \\ 
1
\end{array}
\right)  \label{two}
\end{equation}
The spin states, which we denote by $[\psi _{\pm }],$ are of the same form
and are 
\begin{equation}
\lbrack \psi _{+}]=\left( 
\begin{array}{c}
1 \\ 
0
\end{array}
\right) \text{ and }[\psi _{-}]=\left( 
\begin{array}{c}
0 \\ 
1
\end{array}
\right)  \label{twoa}
\end{equation}
for eigenvalues $+1$ and $-1$ respectively. In the state $[\psi _{+}]$, the
probability of finding the spin to be ''up'' upon measurement is $\left|
a_{+}\right| ^2=1$, while the probability of finding it to be ''down'' is $%
\left| b_{+}\right| ^2=0.$ Similarly, in the state $[\psi _{-}],$ the
probability of finding the spin to be ''up'' upon measurement is $\left|
a_{-}\right| ^2=0$, while the probability of finding it to be ''down'' is $%
\left| b_{-}\right| ^2=1.$

The spin vector is 
\begin{equation}
\lbrack \mathbf{\sigma ]}=\widehat{\mathbf{i}}\left( 
\begin{array}{cc}
0 & 1 \\ 
1 & 0
\end{array}
\right) +\widehat{\mathbf{j}}\left( 
\begin{array}{cc}
0 & -i \\ 
i & 0
\end{array}
\right) +\widehat{\mathbf{k}}\left( 
\begin{array}{cc}
1 & 0 \\ 
0 & -1
\end{array}
\right) .  \label{four}
\end{equation}
Hence the component of spin in the direction of the unit vector $\widehat{%
\mathbf{n}}$ whose polar angles are $(\theta ,\varphi )$ is

\begin{equation}
\lbrack \mathbf{\sigma }_{\widehat{\mathbf{n}}}]=\widehat{\mathbf{n}}\cdot [%
\mathbf{\sigma ]=}\left( 
\begin{array}{cc}
\cos \theta & \sin \theta e^{-i\varphi } \\ 
\sin \theta e^{i\varphi } & -\cos \theta
\end{array}
\right) .  \label{three}
\end{equation}
The eigenvectors of this operator are 
\begin{equation}
\lbrack \xi _{+}]=\left( 
\begin{array}{c}
a_{+} \\ 
b_{+}
\end{array}
\right) =\left( 
\begin{array}{c}
\cos \frac \theta 2 \\ 
\sin \frac \theta 2e^{i\varphi }
\end{array}
\right) \text{ }  \label{fivex}
\end{equation}
and 
\begin{equation}
\text{ }[\xi _{-}]=\left( 
\begin{array}{c}
a_{-} \\ 
b_{-}
\end{array}
\right) =\left( 
\begin{array}{c}
\sin \frac \theta 2 \\ 
-\cos \frac \theta 2e^{i\varphi }
\end{array}
\right)  \label{fivey}
\end{equation}
for eigenvalues $+1$ and $-1$ respectively.

In the literature, Eqns. (\ref{three})-(\ref{fivey}) are given as
representing the most general description of spin $[5]$. We shall therefore
call them the ''standard generalized quantities''. Since the form of the
spin states is the same as the form of the eigenvectors, the ''up'' spin
state is

\begin{equation}
\lbrack \psi _{+}]=\left( 
\begin{array}{c}
\cos \frac \theta 2 \\ 
\sin \frac \theta 2e^{i\varphi }
\end{array}
\right) \text{ }  \label{fiveb}
\end{equation}
while the ''down'' state is 
\begin{equation}
\text{ }[\psi _{-}]=\left( 
\begin{array}{c}
\sin \frac \theta 2 \\ 
-\cos \frac \theta 2e^{i\varphi }
\end{array}
\right) .  \label{fivec}
\end{equation}

If the spin state is $[\psi _{+}],$ then the probability of finding the spin
to be ''up'' upon measurement is $\left| a_{+}\right| ^2=\cos ^2\theta /2$,
while that of finding it to be ''down'' is $\left| b_{+}\right| ^2=\sin
^2\theta /2.$ By the same token, if the spin state is $[\psi _{-}],$ then
the probability of finding the spin to be ''up'' upon measurement is $\left|
a_{-}\right| ^2=\sin ^2\theta /2$, while that of finding it to be ''down''
is $\left| b_{-}\right| ^2=\cos ^2\theta /2.$

The expectation value of the spin projection is

\begin{equation}
\left\langle \sigma _z\right\rangle _{\pm }=[\psi _{\pm }]^{\dagger }[\sigma
_z][\psi _{\pm }].  \label{fived}
\end{equation}
Using the Pauli spin operator and spin states Eqn. (\ref{twoa}), the
expectation value of the spin projection is found to be 
\begin{equation}
\left\langle \sigma _z\right\rangle _{\pm }=\pm 1.  \label{fivee}
\end{equation}
The same results are obtained if the standard generalized operator Eqn. (\ref
{three}) and standard generalized states Eqns. (\ref{fiveb}) and (\ref{fivec}%
) are used in Eqn. (\ref{fived}) to compute this quantity.

In order to obtain more generalized expectation values, we must in Eqn. (\ref
{fived}) use the Pauli spin states, Eqn. (\ref{twoa}), and the standard
generalized operator Eqn. (\ref{three}). Alternatively, we can use the
generalized spin states Eqns. (\ref{fiveb}) and (\ref{fivec}) with the Pauli
operator Eqn. (\ref{one}). In either case, the expectation value is 
\begin{equation}
\left\langle \sigma _z\right\rangle _{\pm }=\pm \cos \theta  \label{fivef}
\end{equation}

These are the standard results for the expectation values. We shall
elucidate them further when we have presented the generalized results.

\section{Generalized Interpretation}

\subsection{Probability Amplitudes and Spin States}

The generalized interpretation is inspired by the approach to quantum
mechanics due to Land\'e $[6-9]$. In this approach, we characterize spin
projection measurements by probability amplitudes describing the
measurements. Let the unit vector $\widehat{\mathbf{a}}$ with polar angles $%
(\theta ,\varphi )$ determine the initial direction of the spin projection,
so that before the current measurement the projection is up or down with
respect to this direction. Let the unit vector $\widehat{\mathbf{c}}$ with
polar angles $(\theta ^{\prime },\varphi ^{\prime })$ be the new direction
in which we seek the spin projection. If the spin projection is initially up
with respect to $\widehat{\mathbf{a}},$a measurement along $\widehat{\mathbf{%
c}}$ will give projections up or down with respect to $\widehat{\mathbf{c}}$
with respective probability amplitudes $\psi (m_i^{(\widehat{\mathbf{a}}%
)};m_f^{(\widehat{\mathbf{c}})})$ where we have labelled the spin states by
their magnetic quantum numbers and the corresponding direction of
quantization. The subscripts $i$ and $f$ refer to initial and final
respectively. The explicit expressions for these probability amplitudes
depend on the choice of phase; there are many such choices $[4]$. For spin
1/2, one choice gives the following $[1,2]$ expressions: 
\begin{equation}
\psi ((+\frac 12)^{(\widehat{\mathbf{a}})};(+\frac 12)^{(\widehat{\mathbf{c}}%
)})=\cos \frac \theta 2\cos \frac{\theta ^{\prime }}2+e^{i(\varphi -\varphi
^{\prime })}\sin \frac \theta 2\sin \frac{\theta ^{\prime }}2,  \label{six}
\end{equation}
\begin{equation}
\psi ((+\frac 12)^{(\widehat{\mathbf{a}})};(-\frac 12)^{(\widehat{\mathbf{c}}%
)})=\cos \frac \theta 2\sin \frac{\theta ^{\prime }}2-e^{i(\varphi -\varphi
^{\prime })}\sin \frac \theta 2\cos \frac{\theta ^{\prime }}2,  \label{seven}
\end{equation}

\begin{equation}
\psi ((-\frac 12)^{(\widehat{\mathbf{a}})};(+\frac 12)^{(\widehat{\mathbf{c}}%
)})=\sin \frac \theta 2\cos \frac{\theta ^{\prime }}2-e^{i(\varphi -\varphi
^{\prime })}\cos \frac \theta 2\sin \frac{\theta ^{\prime }}2  \label{eight}
\end{equation}
and

\begin{equation}
\psi ((-\frac 12)^{(\widehat{\mathbf{a}})};(-\frac 12)^{(\widehat{\mathbf{c}}%
)})=\sin \frac \theta 2\sin \frac{\theta ^{\prime }}2+e^{i(\varphi -\varphi
^{\prime })}\cos \frac \theta 2\cos \frac{\theta ^{\prime }}2.  \label{nine}
\end{equation}

The principle that measurements should be reproducible means that if we
repeat a measurement, the same value should be obtained with certainty. That
means that if we set $\widehat{\mathbf{c}}=\widehat{\mathbf{a}}$, we should
find that 
\begin{equation}
\psi ((\pm \frac 12)^{(\widehat{\mathbf{a}})};(\pm \frac 12)^{(\widehat{%
\mathbf{a}})})=1  \label{ninea}
\end{equation}
and 
\begin{equation}
\psi ((\pm \frac 12)^{(\widehat{\mathbf{a}})};(\mp \frac 12)^{(\widehat{%
\mathbf{a}})})=0.  \label{nineb}
\end{equation}
These probability amplitudes do indeed satisfy these conditions.

According to the Land\'e approach, each one of the probability amplitudes
Eqns. (\ref{six}) - (\ref{nine}) can be expanded according to the example
below for $\psi ((+\frac 12)^{(\widehat{\mathbf{a}})};(+\frac 12)^{(\widehat{%
\mathbf{c}})})$ $[1]$:

\begin{eqnarray}
\psi ((+\frac 12)^{(\widehat{\mathbf{a}})};(+\frac 12)^{(\widehat{\mathbf{c}}%
)}) &=&\chi ((+\frac 12)^{(\widehat{\mathbf{a}})};(+\frac 12)^{(\widehat{%
\mathbf{b}})})\phi ((+\frac 12)^{(\widehat{\mathbf{b}})};(+\frac 12)^{(%
\widehat{\mathbf{c}})})  \nonumber \\
&&+\chi ((+\frac 12)^{(\widehat{\mathbf{a}})};(-\frac 12)^{(\widehat{\mathbf{%
b}})})\phi ((-\frac 12)^{(\widehat{\mathbf{b}})};(+\frac 12)^{(\widehat{%
\mathbf{c}})}).  \label{ten}
\end{eqnarray}
Here $\widehat{\mathbf{b}}$, whose polar angles are $(\theta ^{\prime \prime
},\varphi ^{\prime \prime }),$ is another vector with respect to which we
can measure the spin projection. The $\chi $'s are probability amplitudes
for measurements from the direction $\widehat{\mathbf{a}}$ to the direction $%
\widehat{\mathbf{b}}$ and the $\phi $'s are probability amplitudes for
measurements from the direction $\widehat{\mathbf{b}}$ to the direction $%
\widehat{\mathbf{c}}$. Therefore in the basis $\phi ((+\frac 12)^{(\widehat{%
\mathbf{b}})};(+\frac 12)^{(\widehat{\mathbf{c}})})$ and $\phi ((-\frac
12)^{(\widehat{\mathbf{b}})};(+\frac 12)^{(\widehat{\mathbf{c}})})$, the
probability amplitude $\psi ((+\frac 12)^{(\widehat{\mathbf{a}})};(+\frac
12)^{(\widehat{\mathbf{c}})})$ is represented by the vector

\begin{equation}
\lbrack \psi ((+\frac 12)^{(\widehat{\mathbf{a}})};(+\frac 12)^{(\widehat{%
\mathbf{c}})})]=\left( 
\begin{array}{c}
\chi ((+\frac 12)^{(\widehat{\mathbf{a}})};(+\frac 12)^{(\widehat{\mathbf{b}}%
)}) \\ 
\chi ((+\frac 12)^{(\widehat{\mathbf{a}})};(-\frac 12)^{(\widehat{\mathbf{b}}%
)})
\end{array}
\right) .  \label{el11}
\end{equation}
Using the expressions Eqns. (\ref{six}) - (\ref{nine}) for the probability
amplitudes, we deduce that the generalized spin state $[\psi ((+\frac 12)^{(%
\widehat{\mathbf{a}})};(+\frac 12)^{(\widehat{\mathbf{c}})})]$ has the form
[1]

\begin{eqnarray}
\lbrack \psi ((+\frac 12)^{(\widehat{\mathbf{a}})};(+\frac 12)^{(\widehat{%
\mathbf{c}})})] &=&\left( 
\begin{array}{c}
\cos \frac \theta 2\cos \frac{\theta ^{\prime \prime }}2+e^{i(\varphi
-\varphi ^{\prime \prime })}\sin \frac \theta 2\sin \frac{\theta ^{\prime
\prime }}2 \\ 
\cos \frac \theta 2\sin \frac{\theta ^{\prime \prime }}2-e^{i(\varphi
-\varphi ^{\prime \prime })}\sin \frac \theta 2\cos \frac{\theta ^{\prime
\prime }}2
\end{array}
\right)  \nonumber \\[0.01in]
&=&[\psi (+\frac 12)^{(\widehat{\mathbf{a}})}].  \label{tw12}
\end{eqnarray}

The expansions for the other probability amplitudes show that

\begin{equation}
\lbrack \psi ((+\frac 12)^{(\widehat{\mathbf{a}})};(-\frac 12)^{(\widehat{%
\mathbf{c}})})]=[\psi ((+\frac 12)^{(\widehat{\mathbf{a}})};(+\frac 12)^{(%
\widehat{\mathbf{c}})})]  \label{th13}
\end{equation}
and

\begin{eqnarray}
\lbrack \psi ((-\frac 12)^{(\widehat{\mathbf{a}})};(+\frac 12)^{(\widehat{%
\mathbf{c}})})] &=&[\psi ((-\frac 12)^{(\widehat{\mathbf{a}})};(-\frac 12)^{(%
\widehat{\mathbf{c}})})]  \nonumber \\
&=&\left( 
\begin{array}{c}
\sin \frac \theta 2\cos \frac{\theta ^{\prime \prime }}2-e^{i(\varphi
-\varphi ^{\prime \prime })}\cos \frac \theta 2\sin \frac{\theta ^{\prime
\prime }}2 \\ 
\sin \frac \theta 2\sin \frac{\theta ^{\prime \prime }}2+e^{i(\varphi
-\varphi ^{\prime \prime })}\cos \frac \theta 2\cos \frac{\theta ^{\prime
\prime }}2
\end{array}
\right)  \nonumber \\
&=&[\psi (-\frac 12)^{(\widehat{\mathbf{a}})}].  \label{fo14}
\end{eqnarray}

We observe that owing to the fact that the quantities $\psi ,$ $\chi $ and $%
\phi $ have exactly the same structure, we can as well use one symbol $\chi $
to represent them in the expansion Eqn. (\ref{ten}).

\subsection{Spin Operators}

The most general form of the ''$z$ component'' of the spin operator is $%
[\sigma _{\widehat{\mathbf{c}}}]$, defined with respect to the direction $%
\widehat{\mathbf{c}},$ and is

\begin{equation}
\lbrack \sigma _{\widehat{\mathbf{c}}}]=\left( 
\begin{array}{cc}
(\sigma _{\widehat{\mathbf{c}}})_{11} & (\sigma _{\widehat{\mathbf{c}}})_{12}
\\ 
(\sigma _{\widehat{\mathbf{c}}})_{21} & (\sigma _{\widehat{\mathbf{c}}})_{22}
\end{array}
\right) .  \label{fi15}
\end{equation}

Here, the unit vectors $\widehat{\mathbf{c}}$ and $\widehat{\mathbf{b}}$
have the polar angles $(\theta ^{\prime },\varphi ^{\prime })$ and $(\theta
^{\prime \prime },\varphi ^{\prime \prime })$ respectively. The elements of $%
[\sigma _{\widehat{\mathbf{c}}}]$ are $[1]$ 
\begin{equation}
(\sigma _{\widehat{\mathbf{c}}})_{11}=\cos \theta ^{\prime \prime }\cos
\theta ^{\prime }+\sin \theta ^{\prime \prime }\sin \theta ^{\prime }\cos
(\varphi ^{\prime \prime }-\varphi ^{\prime }),  \label{si16}
\end{equation}

\begin{equation}
(\sigma _{\widehat{\mathbf{c}}})_{12}=\sin \theta ^{\prime \prime }\cos
\theta ^{\prime }-\sin \theta ^{\prime \prime }\cos \theta ^{\prime }-\sin
\theta ^{\prime }[\cos \theta ^{\prime \prime }\cos (\varphi ^{\prime \prime
}-\varphi ^{\prime })+i\sin (\varphi ^{\prime \prime }-\varphi ^{\prime })],
\label{se17}
\end{equation}

\begin{equation}
(\sigma _{\widehat{\mathbf{c}}})_{21}=\sin \theta ^{\prime \prime }\cos
\theta ^{\prime }-\sin \theta ^{\prime \prime }\cos \theta ^{\prime }-\sin
\theta ^{\prime }[\cos \theta ^{\prime \prime }\cos (\varphi ^{\prime \prime
}-\varphi ^{\prime })-i\sin (\varphi ^{\prime \prime }-\varphi ^{\prime })]
\label{ei18}
\end{equation}
and

\begin{equation}
(\sigma _{\widehat{\mathbf{c}}})_{11}=-\cos \theta ^{\prime \prime }\cos
\theta ^{\prime }-\sin \theta ^{\prime \prime }\sin \theta ^{\prime }\cos
(\varphi ^{\prime \prime }-\varphi ^{\prime }).  \label{ni19}
\end{equation}

The eigenvectors of this operator are $[1]$

\begin{equation}
\lbrack \xi _{\widehat{\mathbf{c}}}^{(+1)}]=\left( 
\begin{array}{c}
\chi ((+\frac 12)^{(\widehat{\mathbf{c}})};(+\frac 12)^{(\widehat{\mathbf{b}}%
)}) \\ 
\chi ((+\frac 12)^{(\widehat{\mathbf{c}})};(-\frac 12)^{(\widehat{\mathbf{b}}%
)})
\end{array}
\right) =\left( 
\begin{array}{c}
\cos \frac{\theta ^{\prime }}2\cos \frac{\theta ^{\prime \prime }}%
2+e^{i(\varphi ^{\prime }-\varphi ^{\prime \prime })}\sin \frac{\theta
^{\prime }}2\sin \frac{\theta ^{\prime \prime }}2 \\ 
\cos \frac{\theta ^{\prime }}2\sin \frac{\theta ^{\prime \prime }}%
2-e^{i(\varphi ^{\prime }-\varphi ^{\prime \prime })}\sin \frac{\theta
^{\prime }}2\cos \frac{\theta ^{\prime \prime }}2
\end{array}
\right)  \label{tw20}
\end{equation}
for eigenvalue $+1$ and 
\begin{equation}
\lbrack \xi _{\widehat{\mathbf{c}}}^{(-1)}]=\left( 
\begin{array}{c}
\chi ((-\frac 12)^{(\widehat{\mathbf{c}})};(+\frac 12)^{(\widehat{\mathbf{b}}%
)}) \\ 
\chi ((-\frac 12)^{(\widehat{\mathbf{c}})};(-\frac 12)^{(\widehat{\mathbf{b}}%
)})
\end{array}
\right) =\left( 
\begin{array}{c}
\sin \frac{\theta ^{\prime }}2\cos \frac{\theta ^{\prime \prime }}%
2-e^{i(\varphi ^{\prime }-\varphi ^{\prime \prime })}\cos \frac{\theta
^{\prime }}2\sin \frac{\theta ^{\prime \prime }}2 \\ 
\sin \frac{\theta ^{\prime }}2\sin \frac{\theta ^{\prime \prime }}%
2+e^{i(\varphi ^{\prime }-\varphi ^{\prime \prime })}\cos \frac{\theta
^{\prime }}2\cos \frac{\theta ^{\prime \prime }}2
\end{array}
\right)  \label{tw21}
\end{equation}
for eigenvalue $-1$.

In order to obtain the standard generalized forms Eqns. (\ref{fivex}) and (%
\ref{fivey}) from these formulas, we only need to set $\theta ^{\prime
\prime }=0$, and $\varphi ^{\prime \prime }=\pi .$ Thus, the standard
generalized forms are a special case of the generalized forms.

\subsection{Spin States and Eigenvectors}

The matrix state of a wave function is obtained by first expanding the wave
function in terms of a complete set. When the expansion coefficients are
arranged in a column or row vector, they form the matrix state of the wave
function. This is what we have done to obtain the generalized states and
operators from a probability amplitude basis.

In standard quantum mechanics, the expansion coefficients are viewed as
probability amplitudes which are constants. But if we use the Land\'e
formula $[6-9]$ to perform the expansion, then we straight away recognize
that the expansion coefficients are probability amplitudes with a structure.
This insight allows us to deduce the eigenvectors of the generalized spin
operator without any calculation $[1-4]$.

Spin states and spin eigenvectors have the same structure. Both are each
characterized by two reference directions. In addition, the spin operators
are also characterized by being functions of two reference directions. But
the difference between spin states and spin eigenvectors is this. The
eigenvectors have the same reference directions as the operator they are
eigenfunctions of. But the spin states in the general case have at least one
reference vector different from those characterizing the operator.

\section{Expectation Values}

\subsection{Standard Generalized and Generalized Results}

The standard way of talking about spin concentrates on the current
measurement to the exclusion of the measurement that brought about the spin
state that obtains prior to the current measurement. Thus, this way of
talking implicitly ignores the original direction of the spin projection.
This failure to explicitly mention the precise state that obtains before the
current measurement derives from the belief that in many cases, a quantum
system is in a superposition state before a measurement is made. In such a
state, certain properties of the system are undefined.

In the present formalism, however, the spin state is always well defined.
Probability amplitudes always involve an index or label corresponding to the
state that pertains before the current measurement, and then another
characterizing the state brought about by this measurement. Therefore spin
states are characterized by four angles: one pair of angles defines the
direction of the initial reference vector, while the other pair describes
the direction of the final reference vector. When spin states involve only
two of the angles, as in the standard prescription, then, as we shall
explain later, we are dealing with a special instance of the general case.

\subsection{Generalized Results}

We now proceed to give the general treatment of spin expectation values $[1]$%
. Let the spin projection be initially $m_i$ with respect to $\widehat{%
\mathbf{a}}$. It is then measured with respect to the vector $\widehat{%
\mathbf{c}}$. Its value is found to be $m_f^{(\widehat{\mathbf{c}}\mathbf{)}%
}.$ Here $m_1^{(\widehat{\mathbf{a}}\mathbf{)}}=m_1^{(\widehat{\mathbf{c}}%
\mathbf{)}}=+1$, and $m_2^{(\widehat{\mathbf{a}}\mathbf{)}}=m_2^{(\widehat{%
\mathbf{c}}\mathbf{)}}=-1.$ Since the probability amplitude for obtaining $%
+1 $ is $\psi (m_i^{(\widehat{\mathbf{a}})};(+\frac 12)^{(\widehat{\mathbf{c}%
})})$, and that for obtaining $-1$ is $\psi (m_i^{(\widehat{\mathbf{a}}%
)};(-\frac 12)^{(\widehat{\mathbf{c}})})$, the expectation value of the spin
projection, which we shall denote by $\left\langle \sigma _{\widehat{\mathbf{%
c}}}\right\rangle ,$ is

\begin{eqnarray}
\left\langle \sigma _{\widehat{\mathbf{c}}}\right\rangle &=&\psi ^{*}(m_i^{(%
\widehat{\mathbf{a}})};(+\frac 12)^{(\widehat{\mathbf{c}})})\psi (m_i^{(%
\widehat{\mathbf{a}})};(+\frac 12)^{(\widehat{\mathbf{c}})})(+1)  \nonumber
\\
&&\ +\psi ^{*}(m_i^{(\widehat{\mathbf{a}})};(-\frac 12)^{(\widehat{\mathbf{c}%
})})\psi (m_i^{(\widehat{\mathbf{a}})};(-\frac 12)^{(\widehat{\mathbf{c}}%
)})(-1).  \label{tw22}
\end{eqnarray}

We now expand both $\psi (m_i^{(\widehat{\mathbf{a}})};(+\frac 12)^{(%
\widehat{\mathbf{c}})})$ and $\psi (m_i^{(\widehat{\mathbf{a}})};(-\frac
12)^{(\widehat{\mathbf{c}})})$ as in Eqn. (\ref{ten}). We find that [$1$]

\begin{eqnarray}
\left\langle \sigma _{\widehat{\mathbf{c}}}\right\rangle =[\psi (m_i^{(%
\widehat{\mathbf{a}})})]^{\dagger }\left[ \sigma _z\right] [\psi (m_i^{(%
\widehat{\mathbf{a}})})],  \label{tw25}
\end{eqnarray}
where 
\begin{equation}
\lbrack \psi (m_i^{(\widehat{\mathbf{a}})})]=\left( 
\begin{array}{c}
\chi (m_i^{(\widehat{\mathbf{a}})};(+\frac 12)^{(\widehat{\mathbf{b}})}) \\ 
\chi ((m_i^{(\widehat{\mathbf{a}})};(-\frac 12)^{(\widehat{\mathbf{b}})})
\end{array}
\right)  \label{tw26}
\end{equation}
and the elements of $[\sigma _{\widehat{\mathbf{c}}}]$ are

\begin{equation}
(\sigma _{\widehat{\mathbf{c}}})_{11}=\left| \phi ((+\frac 12)^{(\widehat{%
\mathbf{b}})};(+\frac 12)^{(\widehat{\mathbf{c}})})\right| ^2-\left| \phi
((+\frac 12)^{(\widehat{\mathbf{b}})};(-\frac 12)^{(\widehat{\mathbf{c}}%
)})\right| ^2,  \label{tw29}
\end{equation}

\begin{eqnarray}
(\sigma _{\widehat{\mathbf{c}}})_{12} &=&\phi ^{*}((+\frac 12)^{(\widehat{%
\mathbf{b}})};(+\frac 12)^{(\widehat{\mathbf{c}})})\phi ((-\frac 12)^{(%
\widehat{\mathbf{b}})};(+\frac 12)^{(\widehat{\mathbf{c}})})  \nonumber \\
&&-\phi ^{*}((+\frac 12)^{(\widehat{\mathbf{b}})};(-\frac 12)^{(\widehat{%
\mathbf{c}})})\phi ((-\frac 12)^{(\widehat{\mathbf{b}})};(-\frac 12)^{(%
\widehat{\mathbf{c}})}),  \label{th30}
\end{eqnarray}

\begin{eqnarray}
(\sigma _{\widehat{\mathbf{c}}})_{21} &=&\phi ^{*}((-\frac 12)^{(\widehat{%
\mathbf{b}})};(+\frac 12)^{(\widehat{\mathbf{c}})})\phi ((+\frac 12)^{(%
\widehat{\mathbf{b}})};(+\frac 12)^{(\widehat{\mathbf{c}})})  \nonumber \\
&&-\phi ^{*}((-\frac 12)^{(\widehat{\mathbf{b}})};(-\frac 12)^{(\widehat{%
\mathbf{c}})})\phi ((+\frac 12)^{(\widehat{\mathbf{b}})};(-\frac 12)^{(%
\widehat{\mathbf{c}})})  \label{th31}
\end{eqnarray}
and

\begin{equation}
(\sigma _{\widehat{\mathbf{c}}})_{22}=\left| \phi ((-\frac 12)^{(\widehat{%
\mathbf{b}})};(+\frac 12)^{(\widehat{\mathbf{c}})})\right| ^2-\left| \phi
((-\frac 12)^{(\widehat{\mathbf{b}})};(-\frac 12)^{(\widehat{\mathbf{c}}%
)})\right| ^2.  \label{th32}
\end{equation}

As has already been pointed out, the vector $\widehat{\mathbf{b}}$ is
arbitrary. Therefore, depending on the choice of $\widehat{\mathbf{b}}$, the
state, Eqns. (\ref{tw26}), and the operator, Eqn. (\ref{tw28}) take
different forms. Further variation results from whether $\widehat{\mathbf{c}}
$ equals $\widehat{\mathbf{a}}$ or not. We expect to find among these
various forms the specialized ones that correspond to the standard results.
The full complement of various cases is as given below.

\subsection{\textbf{Summary of Possibilities for the Reference Vectors}}

\subsubsection{\textbf{Case (a)}: $\widehat{\mathbf{b}}\neq \widehat{\mathbf{%
a}}$ and $\widehat{\mathbf{c}}\neq \widehat{\mathbf{a}}$.}

This is the most general case. In this case, the matrix representations are 
\begin{equation}
\lbrack \psi ((+\frac 12)^{(\widehat{\mathbf{a}})};(+\frac 12)^{(\widehat{%
\mathbf{c}})})]=[\psi ((+\frac 12)^{(\widehat{\mathbf{a}})};(-\frac 12)^{(%
\widehat{\mathbf{c}})}))]=\left( 
\begin{array}{c}
\chi ((+\frac 12)^{(\widehat{\mathbf{a}})};(+\frac 12)^{(\widehat{\mathbf{b}}%
)}) \\ 
\chi ((+\frac 12)^{(\widehat{\mathbf{a}})};(-\frac 12)^{(\widehat{\mathbf{b}}%
)})
\end{array}
\right)  \label{th33}
\end{equation}
and 
\begin{equation}
\lbrack \psi ((-\frac 12)^{(\widehat{\mathbf{a}})};(+\frac 12)^{(\widehat{%
\mathbf{c}})}))]=[\psi ((-\frac 12)^{(\widehat{\mathbf{a}})};(-\frac 12)^{(%
\widehat{\mathbf{c}})}))]=\left( 
\begin{array}{c}
\chi ((-\frac 12)^{(\widehat{\mathbf{a}})};(+\frac 12)^{(\widehat{\mathbf{b}}%
)}) \\ 
\chi ((-\frac 12)^{(\widehat{\mathbf{a}})};(-\frac 12)^{(\widehat{\mathbf{b}}%
)})
\end{array}
\right) ,  \label{th34}
\end{equation}
while the operator $\left[ \sigma _z\right] $ has the elements given by
Eqns. (\ref{tw29}) - (\ref{th32}).

\subsubsection{\textbf{Case (b)}: $\widehat{\mathbf{b}}=\widehat{\mathbf{a}}%
. $}

The matrix representations are 
\begin{equation}
\lbrack \psi ((+\frac 12)^{(\widehat{\mathbf{a}})};(\pm \frac 12)^{(\widehat{%
\mathbf{c}})})]=\left( 
\begin{array}{c}
1 \\ 
0
\end{array}
\right)  \label{th35}
\end{equation}
and

\begin{equation}
\lbrack \psi ((-\frac 12)^{(\widehat{\mathbf{a}})};(\pm \frac 12)^{(\widehat{%
\mathbf{c}})})]=\left( 
\begin{array}{c}
0 \\ 
1
\end{array}
\right) ,  \label{th36}
\end{equation}
while the elements of $[\sigma _{\widehat{\mathbf{c}}}]$ are 
\begin{equation}
(\sigma _{\widehat{\mathbf{c}}})_{11}=\left| \phi ((+\frac 12)^{(\widehat{%
\mathbf{a}})};(+\frac 12)^{(\widehat{\mathbf{c}})})\right| ^2-\left| \phi
((+\frac 12)^{(\widehat{\mathbf{a}})};(-\frac 12)^{(\widehat{\mathbf{c}}%
)})\right| ^2,  \label{th37}
\end{equation}

\begin{eqnarray}
(\sigma _{\widehat{\mathbf{c}}})_{12} &=&\phi ^{*}((+\frac 12)^{(\widehat{%
\mathbf{a}})};(+\frac 12)^{(\widehat{\mathbf{c}})})\phi ((-\frac 12)^{(%
\widehat{\mathbf{a}})};(+\frac 12)^{(\widehat{\mathbf{c}})})  \nonumber \\
&&\ -\phi ^{*}((+\frac 12)^{(\widehat{\mathbf{a}})};(-\frac 12)^{(\widehat{%
\mathbf{c}})})\phi ((-\frac 12)^{(\widehat{\mathbf{a}})};(-\frac 12)^{(%
\widehat{\mathbf{c}})}),  \label{th38}
\end{eqnarray}

\begin{eqnarray}
(\sigma _{\widehat{\mathbf{c}}})_{21} &=&\phi ^{*}((-\frac 12)^{(\widehat{%
\mathbf{a}})};(+\frac 12)^{(\widehat{\mathbf{c}})})\phi ((+\frac 12)^{(%
\widehat{\mathbf{a}})};(+\frac 12)^{(\widehat{\mathbf{c}})})  \nonumber \\
&&\ -\phi ^{*}((-\frac 12)^{(\widehat{\mathbf{a}})};(-\frac 12)^{(\widehat{%
\mathbf{c}})})\phi ((+\frac 12)^{(\widehat{\mathbf{a}})};(-\frac 12)^{(%
\widehat{\mathbf{c}})})  \label{th39}
\end{eqnarray}
and

\begin{equation}
(\sigma _{\widehat{\mathbf{c}}})_{22}=\left| \phi ((-\frac 12)^{(\widehat{%
\mathbf{a}})};(+\frac 12)^{(\widehat{\mathbf{c}})})\right| ^2-\left| \phi
((-\frac 12)^{(\widehat{\mathbf{a}})};(-\frac 12)^{(\widehat{\mathbf{c}}%
)})\right| ^2.  \label{fo40}
\end{equation}

\subsubsection{\textbf{Case (c)}: $\widehat{\mathbf{b}}=\widehat{\mathbf{c}}%
. $}

The matrix representations are 
\begin{equation}
\lbrack \psi ((+\frac 12)^{(\widehat{\mathbf{a}})};(+\frac 12)^{(\widehat{%
\mathbf{c}})})]=[\psi ((+\frac 12)^{(\widehat{\mathbf{a}})};(-\frac 12)^{(%
\widehat{\mathbf{c}})})]=\left( 
\begin{array}{c}
\chi ((+\frac 12)^{(\widehat{\mathbf{a}})};(+\frac 12)^{(\widehat{\mathbf{c}}%
)}) \\ 
\chi ((+\frac 12)^{(\widehat{\mathbf{a}})};(-\frac 12)^{(\widehat{\mathbf{c}}%
)})
\end{array}
\right)  \label{fo41}
\end{equation}
and

\begin{equation}
\lbrack \psi ((-\frac 12)^{(\widehat{\mathbf{a}})};(+\frac 12)^{(\widehat{%
\mathbf{c}})})]=[\psi ((-\frac 12)^{(\widehat{\mathbf{a}})};(-\frac 12)^{(%
\widehat{\mathbf{c}})})]=\left( 
\begin{array}{c}
\chi ((-\frac 12)^{(\widehat{\mathbf{a}})};(+\frac 12)^{(\widehat{\mathbf{c}}%
)}) \\ 
\chi ((-\frac 12)^{(\widehat{\mathbf{a}})};(-\frac 12)^{(\widehat{\mathbf{c}}%
)})
\end{array}
\right) ,  \label{fo42}
\end{equation}
while the operator is

\begin{equation}
\lbrack \sigma _{\widehat{\mathbf{c}}}]=\left( 
\begin{array}{cc}
1 & 0 \\ 
0 & -1
\end{array}
\right) .  \label{fo43}
\end{equation}

\subsubsection{\textbf{Case (d)}: $\widehat{\mathbf{c}}=\widehat{\mathbf{a}}$%
.}

The matrix representations are

\begin{equation}
\lbrack \psi ((+\frac 12)^{(\widehat{\mathbf{a}})};(+\frac 12)^{(\widehat{%
\mathbf{a}})})]=[\psi ((+\frac 12)^{(\widehat{\mathbf{a}})};(-\frac 12)^{(%
\widehat{\mathbf{a}})})]=\left( 
\begin{array}{c}
\chi ((+\frac 12)^{(\widehat{\mathbf{a}})};(+\frac 12)^{(\widehat{\mathbf{b}}%
)}) \\ 
\chi ((+\frac 12)^{(\widehat{\mathbf{a}})};(-\frac 12)^{(\widehat{\mathbf{b}}%
)})
\end{array}
\right)  \label{fo44}
\end{equation}
and 
\begin{equation}
\lbrack \psi ((-\frac 12)^{(\widehat{\mathbf{a}})};(+\frac 12)^{(\widehat{%
\mathbf{a}})})]=[\psi ((-\frac 12)^{(\widehat{\mathbf{a}})};(-\frac 12)^{(%
\widehat{\mathbf{a}})})]=\left( 
\begin{array}{c}
\chi ((-\frac 12)^{(\widehat{\mathbf{a}})};(+\frac 12)^{(\widehat{\mathbf{b}}%
)}) \\ 
\chi ((-\frac 12)^{(\widehat{\mathbf{a}})};(-\frac 12)^{(\widehat{\mathbf{b}}%
)})
\end{array}
\right) ,  \label{fo45}
\end{equation}
while the elements of $[\sigma _{\widehat{\mathbf{c}}}]$ are given by Eqns. (%
\ref{tw29}) - (\ref{th32}).

\subsubsection{\textbf{Case (e}): $\widehat{\mathbf{b}}=\widehat{\mathbf{a}}$
and $\widehat{\mathbf{c}}=\widehat{\mathbf{a}}$.}

The matrix representations are

\begin{equation}
\lbrack \psi ((+\frac 12)^{(\widehat{\mathbf{a}})};(+\frac 12)^{(\widehat{%
\mathbf{a}})})]=[\psi ((+\frac 12)^{(\widehat{\mathbf{a}})};(-\frac 12)^{(%
\widehat{\mathbf{a}})})]=\left( 
\begin{array}{c}
1 \\ 
0
\end{array}
\right)  \label{fo46}
\end{equation}
and

\begin{equation}
\lbrack \psi ((-\frac 12)^{(\widehat{\mathbf{a}})};(+\frac 12)^{(\widehat{%
\mathbf{a}})})]=[\psi ((-\frac 12)^{(\widehat{\mathbf{a}})};(-\frac 12)^{(%
\widehat{\mathbf{a}})})]=\left( 
\begin{array}{c}
0 \\ 
1
\end{array}
\right) ,  \label{fo47}
\end{equation}
while the operator is

\begin{equation}
\lbrack \sigma _{\widehat{\mathbf{c}}}]=\left( 
\begin{array}{cc}
1 & 0 \\ 
0 & -1
\end{array}
\right) .  \label{fo48}
\end{equation}

\subsection{Comparison With Standard Results}

As Eqns. (\ref{three}) - (\ref{fivey}) show, the generalized standard forms
of the various quantities explicitly involve only one direction, represented
here by the angles $(\theta ^{\prime },\varphi ^{\prime })$. There would
appear to be no case among the possibilities (a) - (e) corresponding to the
operator being a function of one direction only. But in fact, if we set $%
\theta ^{\prime \prime }=0,$ and $\varphi ^{\prime \prime }$ $=$ $\pi ,$ so
that $\widehat{\mathbf{b}}=\widehat{\mathbf{k}}$ in Case (a), we achieve
this situation. Thus, using the actual expressions for the probability
amplitudes, Eqns. (\ref{tw12}) and (\ref{fo14}), and for the elements of $%
[\sigma _{\widehat{\mathbf{c}}}],$ Eqns. (\ref{si16}) - (\ref{ni19}) we
obtain for Case (a):

\begin{equation}
\lbrack \psi ((+\tfrac 12)^{(\widehat{\mathbf{a}})})]=\left( 
\begin{array}{c}
\cos \frac \theta 2 \\ 
\sin \frac \theta 2e^{i\varphi }
\end{array}
\right) \text{ ,}  \label{fo49}
\end{equation}

\begin{equation}
\lbrack \psi ((-\tfrac 12)^{(\widehat{\mathbf{a}})})]=i\left( 
\begin{array}{c}
\sin \frac \theta 2 \\ 
-\cos \frac \theta 2e^{i\varphi }
\end{array}
\right)  \label{fi50}
\end{equation}
and 
\begin{equation}
\lbrack \sigma _{\widehat{\mathbf{c}}}]\mathbf{=}\left( 
\begin{array}{cc}
\cos \theta ^{\prime } & \sin \theta ^{\prime }e^{-i\varphi ^{\prime }} \\ 
\sin \theta ^{\prime }e^{i\varphi ^{\prime }} & -\cos \theta ^{\prime }
\end{array}
\right) ,  \label{fi51}
\end{equation}

As in this case $\widehat{\mathbf{a}}\neq \widehat{\mathbf{c}}$, using these
quantities ensures that we get the most general expectation value. In view
of this result, it is clear that the standard generalized results are just a
special case of the generalized results. This special form is here obtained
from the generalized form Eqns. (\ref{tw29}) - (\ref{th32})\ by fixing the
angles of $\widehat{\mathbf{b}}$ to $\theta ^{\prime }=0$ and $\varphi
^{\prime }=\pi .$

It is also possible using Case (b) to obtain the standard generalized
expression for the expectation value. In this case, the state is either

\begin{equation}
\lbrack \psi ((+\tfrac 12)^{(\widehat{\mathbf{a}})})]=\left( 
\begin{array}{c}
1 \\ 
0
\end{array}
\right) \text{ ,}  \label{fi52}
\end{equation}
or 
\begin{equation}
\lbrack \psi ((-\tfrac 12)^{(\widehat{\mathbf{a}})})]=\left( 
\begin{array}{c}
0 \\ 
1
\end{array}
\right)  \label{fi53}
\end{equation}
while the operator is given by the elements Eqns. (\ref{si16}) - (\ref{ni19}%
); however, these elements have to be brought to those for the standard
generalized form by setting the angles of $\widehat{\mathbf{b}}=\widehat{%
\mathbf{a}}$ to $\theta =0$ and $\varphi =\pi .$ Then, the operator becomes
identical with Eqn. (\ref{three}), save for the fact that the angles are
double primed..

Similarly, Case (c) can also yield the standard results. We can get the
generalized expectation value by setting the angles of $\widehat{\mathbf{b}}=%
\widehat{\mathbf{c}}$ to $\theta ^{\prime \prime }=0$ and $\varphi ^{\prime
\prime }=\pi $ in the expressions for the states. We then obtain Eqns. (\ref
{fiveb}) and (\ref{fivec}) for the states, save for phase changes. The
operator is the Pauli operator.

What all the three methods of obtaining the standard generalized expectation
value have in common is that the vector $\widehat{\mathbf{b}}$ is set equal
to the unit vector $\widehat{\mathbf{k}}$ which defines the positive $z$
direction. All these methods give the standard generalized expectation
value. However, the form that results from Case (a) is a new form; at any
rate we have not seen it in the literature. The two forms resulting from
Cases (b) and (c) both lead to the results Eqn. (\ref{fivef}). Both these
cases take one of the reference directions for spin projection measurements
as the $z$ axis. Case (b) takes the initial direction $\widehat{\mathbf{a}}$
to equal $\widehat{\mathbf{k}}$, while Case (c) takes the final direction $%
\widehat{\mathbf{c}}$ to equal $\widehat{\mathbf{k}}$. Thus, the standard
treatment of the expectation value completely satisfies the Land\'e-approach
requirement that spin projection measurements must refer to well-defined
initial and final states.

We must mention that we have not seen any evidence that the literature
recognizes that the angular arguments of the spin state and the operator can
differ in the standard generalized expression for the expectation value.
But, considering that the standard description of spin measurements is vague
about the initial state, this is not surprising. Making these arguments
different gives a different final reference direction from the initial, thus
making the expectation value truly general.

For the case where the spin projection is measured with respect to the same
vector twice in succession, the formulas for expectation value are given by
Cases (d) and (e). These cases correspond closely to what is in the
literature. Case (e), of course, involves the standard Pauli quantities.
Both cases lead to the results Eqn. (\ref{fivee}). To correspond this case
to what the literature contains, we again have to make $\widehat{\mathbf{b}}$
correspond to the positive $z$ direction. This is accomplished by setting $%
\theta ^{\prime }=0$ and $\varphi ^{\prime }=\pi .$

If we use the generalized approach to compute the expectation value, we
confirm that 
\begin{equation}
\left\langle \sigma _{\widehat{\mathbf{c}}}\right\rangle =\pm \cos \Theta ,
\label{fi54}
\end{equation}
where $\Theta $ is the angle between $\widehat{\mathbf{a}}$ and $\widehat{%
\mathbf{c}}$, so that

\begin{equation}
\cos \Theta =\widehat{\mathbf{a}}\cdot \widehat{\mathbf{c}}=\cos \theta \cos
\theta ^{\prime }+\sin \theta \sin \theta ^{\prime }\cos (\varphi -\varphi
^{\prime }).  \label{fi55}
\end{equation}

\section{Discussion and Conclusion}

In this paper, we have explained how the new generalized spin quantities we
have introduced relate to the usual quantities which we have called the
''standard generalized quantities''. We have shown how despite the seeming
difference in the two kinds of quantities, the ''standard generalized
quantities'' are special forms of the new quantities.

Our success in obtaining the spin quantities from first principles [$1-4$],
and in using them to elucidate standard ideas is strong indication that the
Land\'e approach to quantum mechanics could be fruitful if pursued logically
and applied even to seemingly well-understood results. The idea of attaching
two labels to a probability amplitude is one which seems to be of general
validity. This idea is applicable to eigenfunctions. These are just
probability amplitudes corresponding to the measurement of an observable
with a continuous eigenvalue spectrum. For an eigenfunction, the
characteristic eigenvalue defines the state that obtains before the current
measurement.

It is our belief that many more interesting results and elucidations of
standard quantum theory will be obtained by means of the Land\'e approach.

\section{References}

\textbf{1}. Mweene H. V., ''Derivation of Spin Vectors and Operators From
First Principles'', submitted to \textit{Foundations of Physics, }%
quant-ph/9905012

\textbf{2}. Mweene H. V., ''Generalized Spin-1/2 Operators and Their
Eigenvectors'', quant-ph/9906002

\textbf{3}. Mweene H. V., ''Vectors and Operators For Spin 1 Derived From
First Principles'', quant-ph/9906043

\textbf{4}. Mweene H. V., ''Alternative Forms of Generalized Vectors and
Operators for Spin 1/2'', quant-ph/9907031

\textbf{5}. Bransden and Joachain, ''Introduction to Quantum Mechanics'',
Longman Scientific \& Technical, 1989.

\textbf{6}. Land\'e A., ''From Dualism To Unity in Quantum Physics'',
Cambridge University Press, 1960.

\textbf{7}. Land\'e A., ''New Foundations of Quantum Mechanics'', Cambridge
University Press, 1965.

\textbf{8}. Land\'e A., ''Foundations of Quantum Theory,'' Yale University
Press, 1955.

\textbf{9}. Land\'e A., ''Quantum Mechanics in a New Key,'' Exposition
Press, 1973.

\end{document}